# Watermarking Digital Images Based on a Content Based Image Retrieval Technique

Dimitrios K. Tsolis[1], Spyros Sioutas[2] and Theodore S. Papatheodorou[1]

[1] *Department of Computer Engineering and Informatics, University of Patras, Greece*

[2] *Department of Informatics, Ionian University, Corfu, Greece*

+302610996900

+302610969001

dkt@hpclab.ceid.upatras.gr

http://www.hpclab.ceid.upatras.gr

The current work is focusing on the implementation of a robust watermarking algorithm for digital images, which is based on an innovative spread spectrum analysis algorithm for watermark embedding and on a content-based image retrieval technique for watermark detection. The highly robust watermark algorithms are applying "detectable watermarks" for which a detection mechanism checks if the watermark exists or no (a Boolean decision) based on a watermarking key. The problem is that the detection of a watermark in a digital image library containing thousands of images means that the watermark detection algorithm is necessary to apply all the keys to the digital images. This application is non-efficient for very large image databases. On the other hand "readable" watermarks may prove weaker but easier to detect as only the detection mechanism is required. The proposed watermarking algorithm combine's the advantages of both "detectable" and "readable" watermarks. The result is a fast and robust watermarking algorithm.

## Introduction

Advances in technology have improved the ability to reproduce, distribute, manage and publish information (CSTB, 99). Reproduction costs are much lower for both rights holders (content owners) and infringers while digital copies are perfect replicas. In addition, the computer networks have changed the economics of distribution. Networks enable sending multimedia content worldwide, cheaply and at a high speed. As a consequence, it is easier and less expensive both for a rights holder to distribute a work and for an individual to make and distribute unauthorized copies. Finally, the World Wide Web has altered at a fundamental way the publication of information, allowing everyone to be a publisher with worldwide reach.

Wide access and delivery of valuable content raise several critical issues, pertaining to management, protection and exploitation of digitized content. These include the critical problem of IPR (Intellectual Property Rights), protection and the unauthorized use and exploitation of digital data. Besides economical and other implications, such problems create considerable skepticism to organizations and individual content owners. As a result content of great educational and economical value is often held secret and private (House 98).

Technological protection means is one of the key components, attracting plenty of scientific research, within the generalized Digital Rights Management Systems framework. Watermarking is probably the most promising technological approach against Intellectual Property Rights violations. The current work is focusing on the implementation of a robust watermarking algorithm for digital images, which is based on an innovative spread spectrum analysis algorithm for watermark embeddding and on a content based image retrieval technique for watermark detection.

## Detectable and Readable Watermarks

In this section a brief overview of detectable and readable watermarks is being presented aiming at defining the most important advantages and disadvantages of each case.

The watermarking algorithms are applying watermarks (invisible information in bitstreams) to digital images. The process of watermark embedding is using a watermarking key and the watermarking algorithm, to produce the watermarked digital image. The embedding method vary based on which image domain is being processed, e.g. the space, frequency domain or the wavelets. Depending on the embedding method detectable (single-bit) or readable (multi-bit) watermarks are being incorporated to the digital images.

The watermark detection is using usually the same watermarking key and the reverse embedding method to detect if a watermark exists or no in the digital image (in the detectable case) or to read the information incorporated into the digital image (in the readable case).

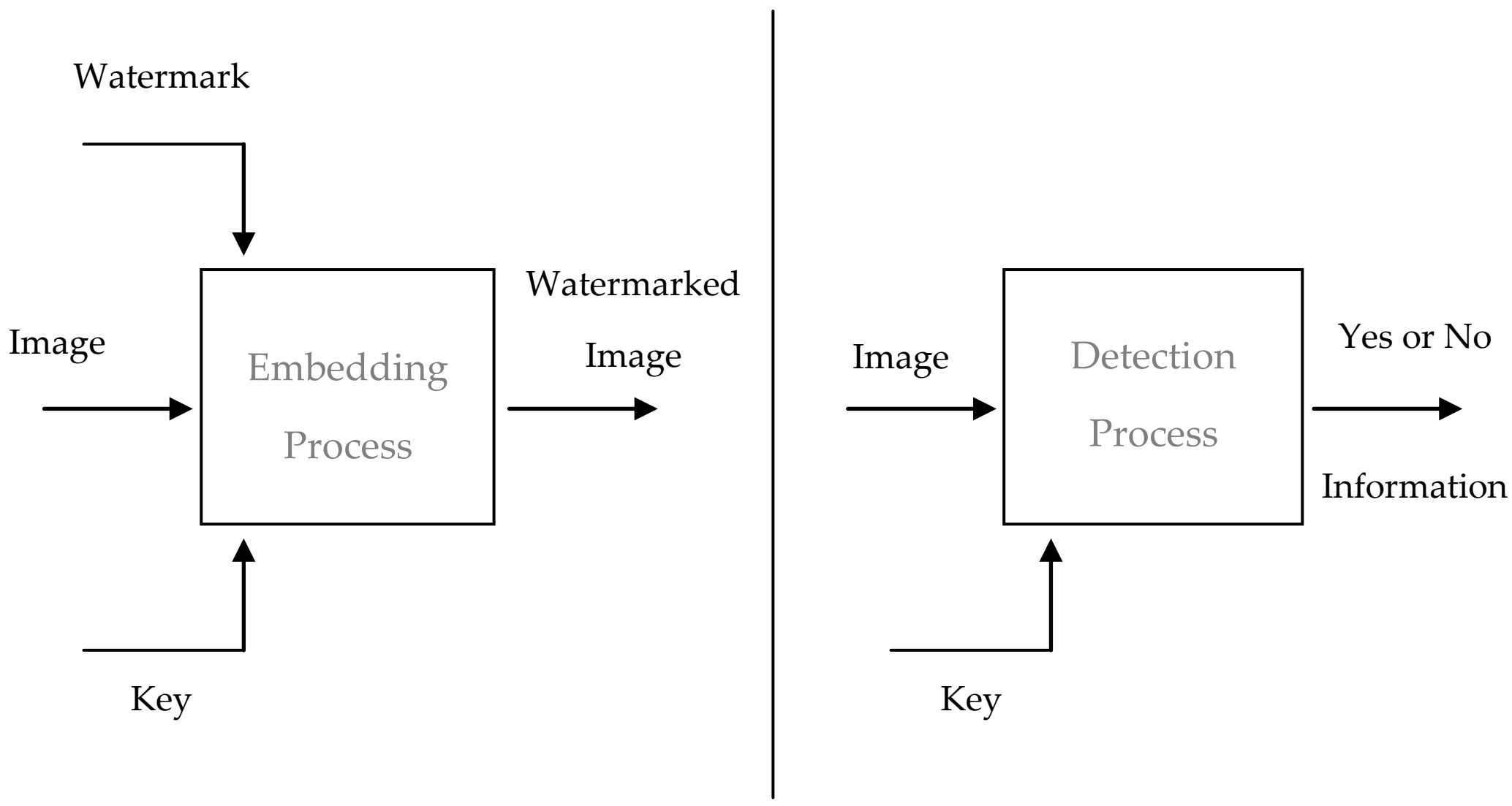

The main advantage of the detectable watermarks is unbeatable robustness to the most common attacks such as digital image compression, geometric transformations (e.g. rotation and scaling), Gaussian filters, linear transformation, aspect ratio, shearing, signal enhancement, row-column removal and the analog to digital attack. The main disadvantage of the detectable case is that it does not reflect to the needs of real life applications. For example, in the case of a very large digital image library with thousands of images it is necessary to assign one watermarking key per digital image so as to ensure unique identification for each digital watermark. This application is very important especially if the digital images will be commercially exploited through the Web. Consequently an equal number of watermarking keys is necessary before starting the embedding process. After the completion of the watermarking process lets assume that the digital images are being commercially distributed and a sold digital image has been published to a Web site or used in a DVD title. If the owner of the digital image library would like to prove that he owns the digital images he should acquire the digital image from the Web site or DVD and apply the detection process. The detection process requires the correct watermarking key for the specific image so as to have a Yes or No result. The first solution is to test all the watermarking keys through the detection process for the specific image, but this is practically impossible because a thorough detection process has an average of 30 seconds duration and the overall detection process could last many days. The alternative is to find a way to limit down the number of the watermarking keys used by the detection process.

Most of times in real life applications “readable” watermarks are being used. In this case the detection process uses only one watermarking key and reads the invisible information embedded to the image through watermarking. This information could be a number, a text e.g. “Copyright – Owner – 2008” etc. The advantage is that the detection process is quick and easy. The disadvantage is that “readable” watermarks are not as robust as the “detectable” watermarks and a “readable” watermark is a multi-bit stream which may deteriorate the quality of the watermarked image.

So in an ideal application watermarks which have the robustness of a “detectable” watermark and easy and quick detection process like the “readable” watermarks should exist.

## Proposed Watermarking Algorithm

Watermarking principles are mainly used whenever copyright protection of digital content is required and the cover-data is available to parties who are aware of the existence of the hidden data and may have an interest removing it [Cox, 02]. In this framework the most popular and demanding application of watermarking is to give proof of ownership of digital data by embedding copyright statements. For this kind of application the embedded information should be robust against manipulations that may attempt to remove it. At the same time the detection process should be quick and easy so as to support real life applications to work properly and efficiently (Randall 01).

According to the above the first most important step towards the implementation of the watermarking algorithm is the selection and evaluation of the watermarking method. The method chosen is a spread spectrum multibit watermarking technique.

### Watermark Embedding

The embedding of a robust multibit watermark is accomplished through casting several zero-bit watermarks onto specified coefficients. The image watermark, a random sequence of Gaussian distribution in our case, is casted multiple times onto the selected coefficients preserving the same sequence length but shifting the start point of casting by one place. Actually the final watermark that will be embedded into the image is not a single sequence but many different sequences generated with different seeds. These sequences will be casted, one after the other, on the mid coefficients of the image, using the additive rule mentioned above and begging from successive starting points. If all sequences where to be casted, beginning from the same starting point, then, besides the severe robustness reduction resulting from the weak correlation, the possibility of false positive detector response would dramatically increase, since every number that has participated as a seed during the sequence generation procedure, will be estimated by the detector as a valid watermark key. Shifting the starting point by one degree for every sequence casting ensures that the false positive rate will remain in very small level due to the artificial desynchronisation introduced. Every single random sequence of Gaussian distribution isgenerated using a different number as the seed for the Gaussian sequence generator. It is important to differentiate the sequences in order not to mislead the detection mechanism, since it is based on the correlation between the extracted sequence and the sequence produced with the watermark key.

The watermark key is responsible both for the generation of the first sequence and the construction of a vector, containing the rest of the numbers that will serve as the corresponding seeds. The placement of several Gaussian sequences into the image content can model, under specific conventions, a multibit watermark. The detection of a zero-bit watermark is interpreted as if the bit value of the specified bit is set to one. On the contrary, failure of the detector to detect the zero-bit watermark leads to the conclusion of a zero bit value. Thus, in order for a message to be casted into the image content, it is initially encoded using the binary system and applied afterwards in the sense of zero-bit watermarks using the embedding mechanism and according to the derived bit sequence.

In addition one of the most important aspects of the algorithm is its ability to preprocess the digital image and preserve some valuable supplementary information in a database (which could be a digital image library of the owner). The association between the digital image and the supplementary information is accomplished through an integer value called imageid, acting as a foreign key for the supplementary information database table. The watermarking algorithm has the responsibility of connecting to the database and storing the supplementary information, acquired from the image preprocessing, always associated with the actual image through the imageid. The best way to demonstrate how the embedding method works is to present the next two figures. The special reference on this particular algorithm's function is justified by the significance of the supplementary information to the process of image registration and consequently, robustness against geometrical attacks during the detection process.

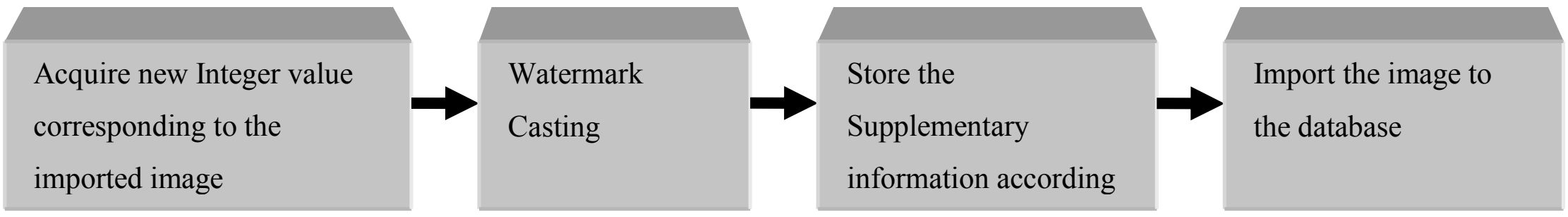

Fig. Watermark Embedding Process

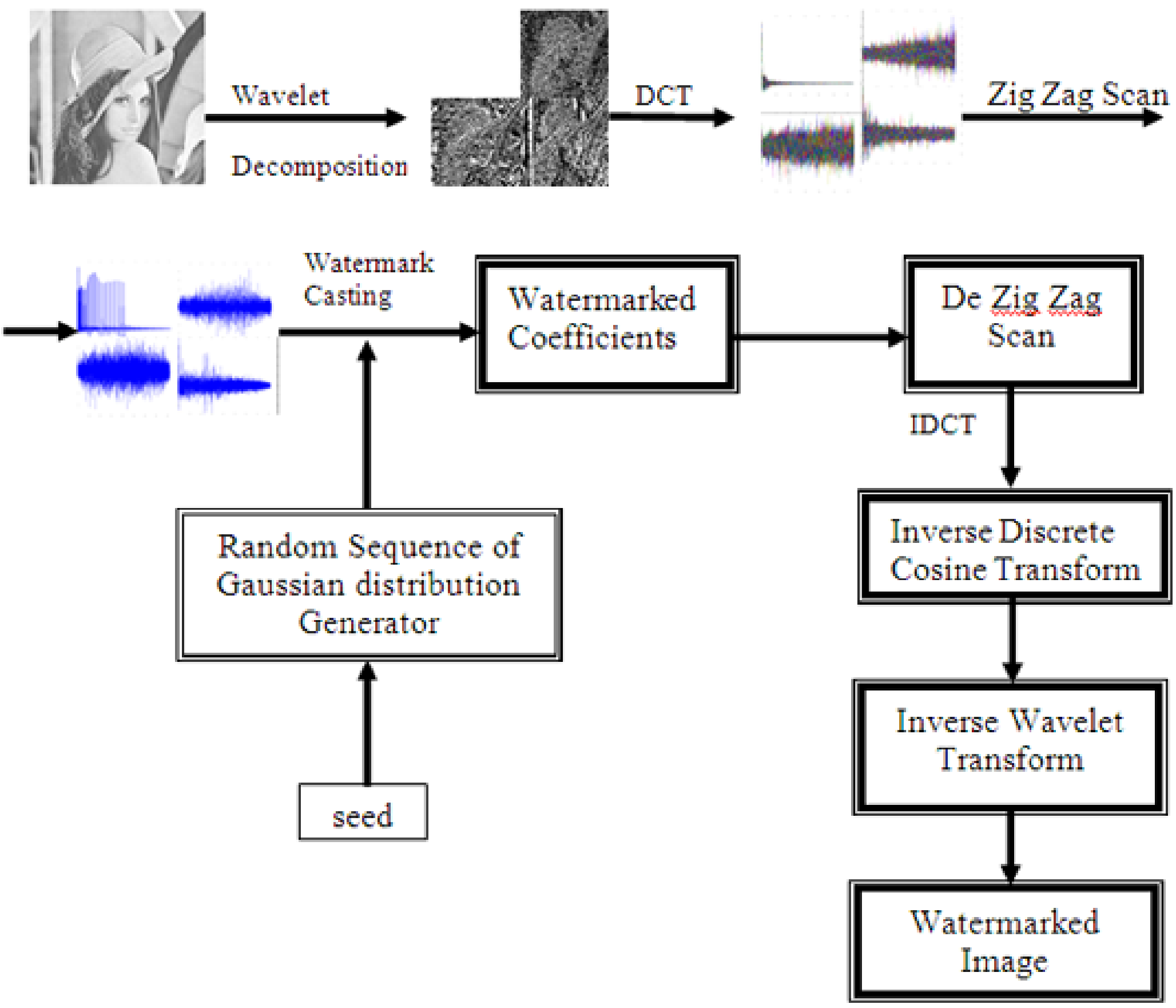

Fig. Watermark Casting

Some important remarks regarding the novelty of the proposed schema are addressed below.

Data payload: The reason that most of the proposed robust watermarking systems are zero-bit, is highly related to the data payload. Data payload is the amount of information encoded into the image during the watermark procedure. In other words, it is the number of coefficients modified according to the additive rule. The performance of the correlation function adopted by the detector is increased when a strong statistical dependency is present. On the other hand, the statistical dependency requires a significant sequence length in order to fulfill the requirements of the correlation function. In addition, the position and the amount of coefficients modified, affects directly the resulting image quality. This is one of the most important tradeoffs that the designer of a watermarking system has to balance.

Casting multiple sequences will maximize the problem of image distortion. In that sense, the maximum number of bits allowed for encoding the watermark message is crucial. In the proposed scheme a total number of 16 bits were selected. The first bit indicates the existence of a watermark. If the response is positive the detector continues with the following zero-bit watermarks, otherwise the mechanism outputs a negative response. This is a useful shortcut saving the detector of valuable time and processing power. The second bit serves as a flag important for the decoding operation. The role of this bit flag is described in detail in the following paragraph. The next 14 bits are dedicated to the encoding of the watermark message. Under the aforementioned conventions the system is capable of embedding 214 different messages.

Seed Vector Generation: The watermark key is a positive integer value playing a vital role in the overall watermarking procedure. It corresponds to the private information that must be shared between the embedder and the detector of the watermark. One of the basic principles of private watermarking is that the encryption of the information to be embedded is performed according to a private key. Thus, if an image is watermarked using a specified key, it is impossible for the detector to detect the watermark unless provided with the same key. The encryption is accomplished by using the private key as the seed for the pseudorandom sequence of Gaussian distribution generator. In our case, there is the necessity of 15 extra numbers, one for each sequence. Thus, the private key except from its basic operation as a pseudorandom generator seed is also used as the seed for producing a vector containing 15 numbers. It is important for every private key to produce a different vector of numbers, in order to avoid undesirable statistical dependencies between different watermarks. A pseudorandom generator provided by any compiler is capable of applying this one-way relationship between the private key and the produced vector of numbers.

Flag bit operation: Under the convention, that for every one-bit-value we cast a zero- bit watermark and for every zero-bit-value we dont do anything except moving to the next starting point, the number of zero-bit watermarks to be casted is dictated by the bit sequence. It is obvious that a bit sequence containing only a single one-bit-value is preferable from a sequence consisted of 14 aces. Both for, processing power and watermarks imperceptibility purposes, a bit reversal trick is required for optimizing the embedders performance. Thus, after acquiring the binary representation of the message, a counter scans the bit sequence counting the zeros and the aces. If the number of aces is grater than the number of zeros a bit reversed sequence is generated. The zero-bit watermarks casting is now performed according to the newly generated sequence. In that case, the flag bit is set to one serving as an indicator to the detector that the extracted sequence is bit-reversed. As a consequence, the decoder, equipped with the appropriate information, can easily decode a message represented by 14 aces binary sequence, even though the embedder had casted only two zero-bit watermarks. The benefit of using the specified trick is that even though a 16-bit watermark is supported, we only need to cast 8 zero-bits watermarks in the worst case.

## Watermark Detection with Content Based Image Retrieval

Many watermarking schemes show weaknesses in a number of attacks and specifically those causing desynchronization which is a very efficient tool against most marking techniques [Katzenbeisser, 00]. This leads to the suggestion that detection, rather than embedding, is the core problem of digital watermarking [Wayner, 02]. As a consequence the main weakness of the majority of the watermarking detection mechanisms is there inability to counter the attacks involving the desynchronization of the detector. Geometrical attacks are a small but important subset of this kind of attacks. The best countermeasure against the desychronization attacks is definitely the notion of “image registration”.
Image registration is the procedure of finding the exact image instance during the watermark casting. Finding the right instance and providing it to the detector helps the mechanism to achieve synchronization and detect the watermark. However, finding the appropriate instance, without further information available, is not a straightforward task. In the trivial case, the necessary additional information is the original image. This is mainly the reason why the non blind detectors (the original image is in the detector’s disposal) perform better than blind detectors (only the watermarked image is available).
The presence of a digital image library and specifically of a DBMS (Database Management System) with advanced search capabilities provided the basis for a more efficient detection mechanism through the cooperation of the image database with the watermarking technique. The detector is initially provided with a digital image in order to decide whether it is watermarked or not. If the first attempt to find the watermark is unsuccessful the detector must try to register the image hoping to find its synchronization and detect the watermark. At this point the original image is essential for the detector. Although, our algorithm may have access to a large number of digital images stored in a database, it is impossible to decide which image corresponds to the original copy of the image in the detector. This is where the advanced search capabilities take over and in particular a Content Based Image Retrieval algorithm.
The algorithm used is the QBIC (Query by Image Content) algorithm (Holt 02) which is a tool that allows the storage of and query of image data with the same convenience as with traditional ones. The prominent feature of the QBIC is the functionality of querying images, based on related business data or by image attributes. The entire image database search can be based on data that the user maintains, such as name, number and description, or by data that the QBIC maintains, such as the format of the image, its distribution of colours, the illustrated shapes etc. The QBIC queries is the solution to the problem of selecting the correct original image.
Just before the initialisation of the detection process a QBIC query is constructed based on the image under examination. The query response is a similarity measure reporting the probability that the original copy of the image under examination is the one indicated by QBIC query. If the probability is high enough, the detector continues the detection procedure having access not only to the original image but also to the supplementary information derived from the image pre-processing. The association between the original image and the supplementary information is conducted through an integer value returned by the QBIC query, which corresponds to the foreign key of the corresponding database table. The following figure demonstrates the possible scenarios.

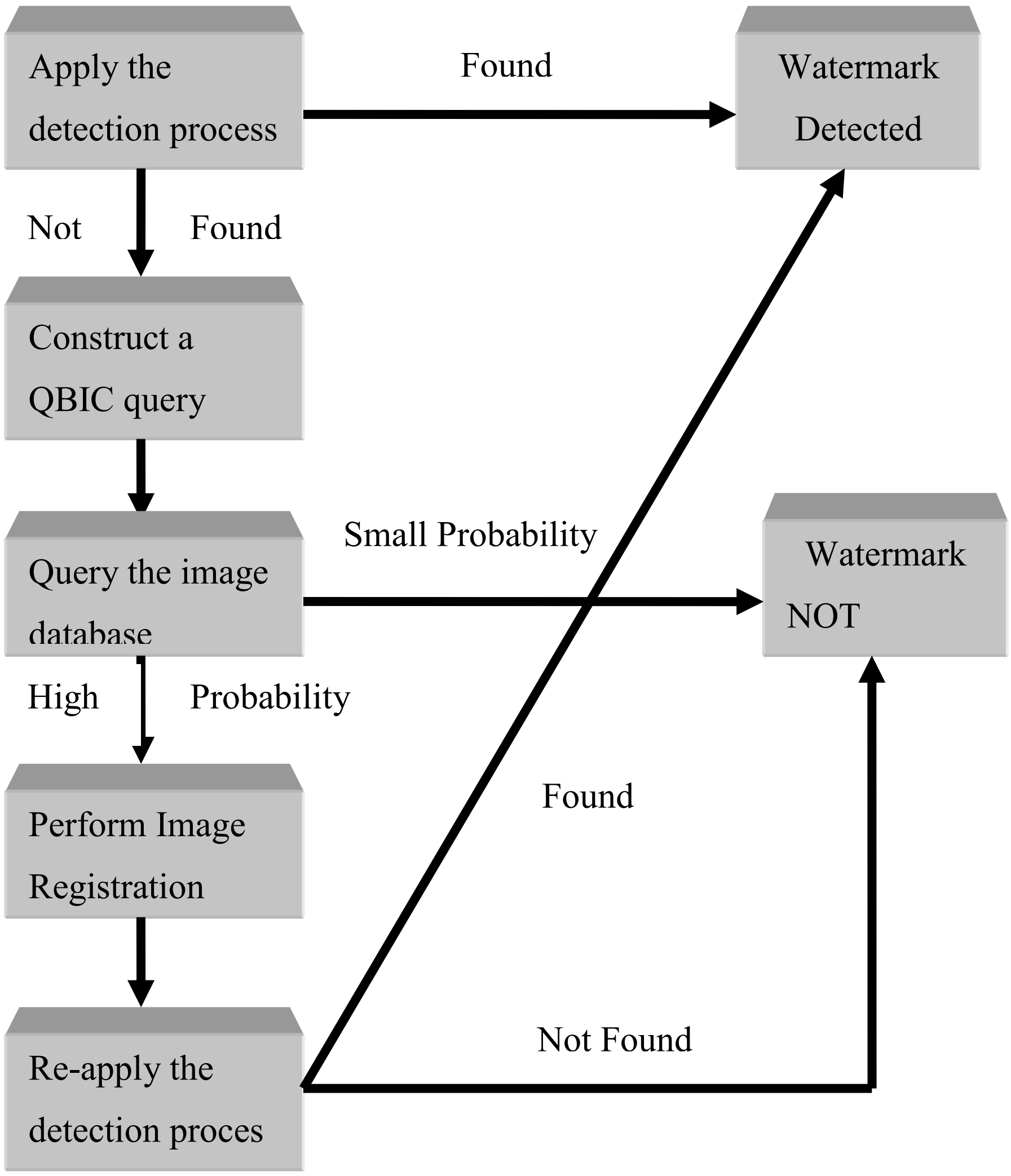

Fig. Watermark Detection Process

### Evaluation and Robustness

In this section we present the experimental results concerning the evaluation and robustness of the watermarking algorithm. Robustness is the most highly desired feature of a watermarking algorithm especially if the application demands copyright protection, and persistent owner identification. In addition the image distortion and false positive parameters are being evaluated.

In our experiments the metric selected for evaluating the image distortion introduced by the multi-bit watermark casting is PSNR (Pick Signal to Noise Ration). Although PSNR is definitely insufficient for modeling the complexity of the human visual system is by all means an effective metric for measuring image similarity. The following table demonstrates the results.

| **Image Database** | | | | |
|---|---|---|---|---|
| **Original Images** | | | | |
| 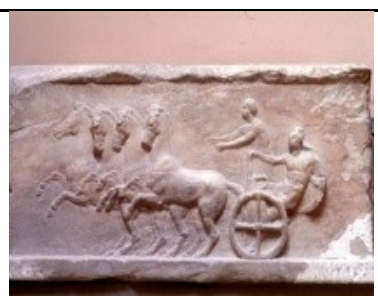 | 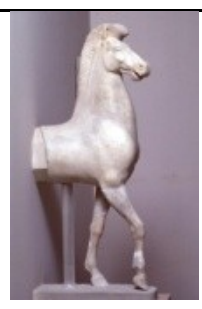 | 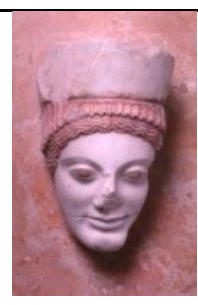 | 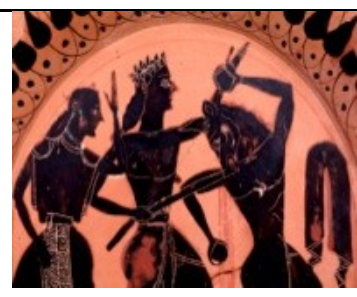 | 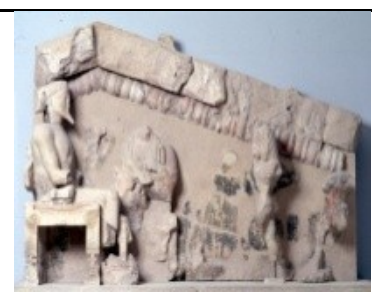 |
| Chariot | Horse | Mask | Plate | Scene |
| **Watermarked Images** | | | | |
| 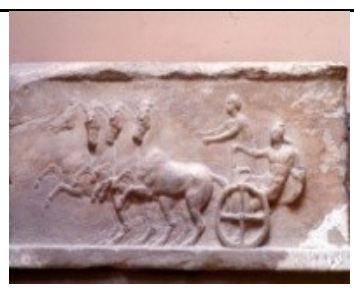 | 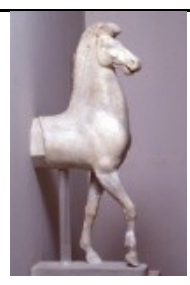 | 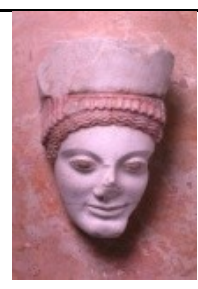 | 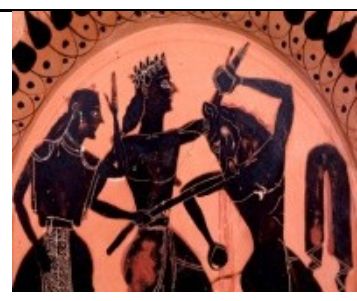 | 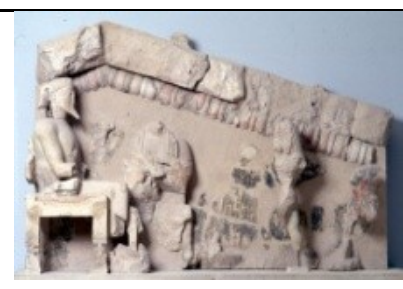 |
| water_chariot | water_horse | water_mask | water_plate | water_scene |
| **PSNR** | **PSNR** | **PSNR** | **PSNR** | **PSNR** |
| 66.65 | 69.52 | 64.90 | 68.36 | 65.99 |

Table 1. Image Database – PSNR

Regarding the fact that in most cases a PSNR value above 40 decibel is satisfactory the derived results can be consider to meet the image quality requirements.

Casting multiple zero-bit watermarks onto the same coefficient area raises the probability of causing abnormal fluctuation of the detector's false positive probability. In order to confirm that no such case is true, we used 5 different watermarks applied to a sample of 5 images for approximating the false positive probability. The watermarks were generated from 5 different integer numbers, also responsible for the generation of the vector containing the rest integer values required by the embedding mechanism. Every image was watermarked using each of this numbers as a watermark key while afterwards the detector was tested for possible false positive response with every number contained in the produced vector. That is, an image watermarked with the number K1 as a watermark key was examined by the detector 15 more times using as primary keys the numbers of the vector produced by the random generator with K1 as a seed. The reason for examining only this small subset of numbers instead of a large random set is that this numbers hold highest probability of causing a false positive, due to the statistical dependence introduced to the correlation function. The next diagram demonstrates the experimental results:

| **Keys** | | | | | **Chariot** | | | | | **Horse** | | | | | **Mask** | | | | | **Plate** | | | | |
|---|---|---|---|---|---|---|---|---|---|---|---|---|---|---|---|---|---|---|---|---|---|---|---|---|
| **50** | **100** | **200** | **350** | **700** | 1 | 1 | 1 | 1 | 1 | 1 | 1 | 1 | 1 | 1 | 1 | 1 | 1 | 1 | 1 | 1 | 1 | 1 | 1 | 1 |
| 22715 | 12662 | 25325 | 27935 | 23102 | 0 | 0 | 0 | 0 | 0 | 0 | 0 | 0 | 0 | 0 | 0 | 0 | 0 | 0 | 0 | 0 | 0 | 0 | 0 | 0 |
| 22430 | 23392 | 25316 | 28203 | 2170 | 0 | 0 | 0 | 0 | 0 | 0 | 0 | 0 | 0 | 0 | 0 | 0 | 0 | 0 | 0 | 0 | 0 | 0 | 0 | 0 |
| 16275 | 22561 | 2367 | 21228 | 32468 | 0 | 0 | 0 | 0 | 0 | 0 | 0 | 0 | 0 | 0 | 0 | 0 | 0 | 0 | 0 | 0 | 0 | 0 | 0 | 0 |
| 21417 | 20718 | 19320 | 17222 | 12328 | 0 | 0 | 0 | 0 | 0 | 0 | 0 | 0 | 0 | 0 | 0 | 0 | 0 | 0 | 0 | 0 | 0 | 0 | 0 | 0 |
| 4906 | 6314 | 9131 | 13355 | 23212 | 0 | 0 | 0 | 0 | 0 | 0 | 0 | 0 | 0 | 0 | 0 | 0 | 0 | 0 | 0 | 0 | 0 | 0 | **1** | 0 |
| 9000 | 1073 | 17987 | 26975 | 4255 | 0 | 0 | 0 | 0 | 0 | 0 | 0 | 0 | 0 | 0 | 0 | 0 | 0 | 0 | 0 | 0 | 0 | 0 | 0 | 0 |
| 3863 | 24449 | 86 | 29076 | 9340 | 0 | 0 | 0 | 0 | 0 | 0 | 0 | 0 | 0 | 0 | 0 | 0 | 0 | 0 | 0 | 0 | 0 | 0 | 0 | 0 |
| 26227 | 32712 | 12916 | 32372 | 12235 | 0 | 0 | 0 | 0 | 0 | 0 | 0 | 0 | 0 | 0 | 0 | 0 | 0 | 0 | 0 | 0 | 0 | 0 | 0 | 0 |
| 20017 | 27912 | 10934 | 1851 | 24347 | 0 | 0 | 0 | 0 | 0 | 0 | 0 | 0 | 0 | 0 | 0 | 0 | 0 | 0 | 0 | 0 | 0 | 0 | 0 | 0 |
| 21604 | 2031 | 28420 | 2467 | 29292 | 0 | 0 | 0 | 0 | 0 | 0 | 0 | 0 | 0 | 0 | 0 | 0 | 0 | 0 | 0 | 0 | 0 | 0 | 0 | 0 |
| 28180 | 11332 | 10403 | 25394 | 5760 | 0 | 0 | 0 | 0 | 0 | 0 | 0 | 0 | 0 | 0 | 0 | 0 | 0 | 0 | 0 | 0 | 0 | 0 | 0 | 0 |
| 940 | 7595 | 20906 | 8104 | 21924 | 0 | 0 | 0 | 0 | 0 | 0 | 0 | 0 | 0 | 0 | 0 | 0 | 0 | 0 | 0 | 0 | 0 | 0 | 0 | 0 |
| 13042 | 20424 | 2421 | 24568 | 10709 | 0 | 0 | 0 | 0 | 0 | 0 | 0 | 0 | 0 | 0 | 0 | 0 | 0 | 0 | 0 | 0 | 0 | 0 | 0 | 0 |
| 26566 | 10606 | 11456 | 29111 | 15696 | 0 | 0 | 0 | 0 | 0 | 0 | 0 | 0 | 0 | 0 | 0 | 0 | 0 | 0 | 0 | 0 | 0 | 0 | 0 | 0 |
| 20934 | 20780 | 20474 | 20015 | 18943 | 0 | 0 | 0 | 0 | 0 | 0 | 0 | 0 | 0 | 0 | 0 | 0 | 0 | 0 | 0 | 0 | 0 | 0 | 0 | 0 |

The above diagram indicates only one false positive response under the “Plate” image. Thus, the derived conclusions justify our hypothesis about the false positive probability of the detector which remains in relatively low values, thanks to the statistical independence introduced by the embedding start point shifting. The watermark’s robustness depends on the efficiency of Image Extenders which were analysed in the section above. The watermarks robustness has been extensively tested. The average score of the watermarking robustness against various types of attacks is 94% which is a very efficient result for the type of application under consideration. The results are briefly analyzed below:

| **Type of Attack** | **Average Score** |
|---|---|
| Convolution and Median Filters | 100% |
| Jpeg Compression | 90% |
| Scaling | 100% |
| Cropping | 95% |
| Shearing | 93% |
| Rotation – Crop | 97,5% |
| Rotation – Crop – Scale | 79% |
| Linear Transformations | 100% |
| Aspect Ratio | 100% |
| Row and Column Removal | 100% |
| Geometric Distortion | 80% |

Table 2. Watermarking Robustness – Various Attacks

Closing the performance evaluation it is worth mentioning the results derived from the print-scan or digital to analog attack. A small number of images after they have been compressed with a jpeg algorithm, they were printed to plain paper. Using a flatbed scanner the images were scanned back to their digital form and delivered to the watermark detector. The detector output is presented in the following table.

| Image Format | Image Compression | Print Quality | Result |
|---|---|---|---|
| Tiff | None | Best | Detected |
| Tiff | None | Normal | Detected |
| Jpeg | Medium Compression High Quality | Best | Detected |
| Jpeg | Medium Compression High Quality | Normal | Detected |
| Jpeg | Medium Compression Medium Quality | Best | Detected |
| Jpeg | Medium Compression Medium Quality | Normal | Missed |

Table 3: Print – Scan or Digital Analog Attack

The watermarking robustness is more than adequate. The JPEG compression, crop and rotation attacks which are the more common types of attack for applications which distribute digital multimedia through the web is being dealt effectively.

## Copyright Protection with the Watermarking Algorithm

Securing the digital content is of a great concern. The reasonable approach would be to adopt a strategy of securing the content by guarding it. By guarding we mean the establishment of complicated mechanisms difficult to overcome without proper authorization. Encryption and user authentication are some of the techniques used to forbid access of the digital content. Nevertheless, in circumstances where the adversary succeeds in circumventing the guarding mechanisms, the content is totally unprotected and vulnerable to illegal manipulation. On the contrary, the security provided by watermarking techniques relies on the content itself. Thus, protection continues even after the adversary has managed to obtain a digital image library's content. The watermarking algorithm, which was described, is used to facilitate important security tasks over the content. The main tasks are copyright protection by copy control and owner identification, digital signature and transaction tracking.

The enforcement of the aforementioned security measures is based on the notion of the watermark key. The usage and administration of the watermark key is what differentiate the form of security applied, resulting in different cases. The basic principle of every watermarking scheme is that in order for the detection to be successful, the key used by the detector should match the one used by the embedding mechanism. The selection of any different key must cause the detector to fail. An important detail concerning the detector's output is the value returned. In the trivial case the returned value is a simple indication deciding for the watermark's existence (Yes / No Boolean response). Under different circumstances it is useful for the detector to return an integer value. This value will serve as a pointer to a useful piece of information regarding the digital multimedia.

The copyright protection scenario is the most important one. This is based on two cases, the owner identification and copy control.

In the owner identification case the image owner casts a watermark to the image using a private key. The scenario begins with a dispute between the image owner and an adversary. They both claim ownership of the digital image and they are both asked to give proof of their assertion. The copyright owner with the correct key value in his disposal can prove his assertion by feeding the key to the detector and confirming the watermark's presence. On the contrary, the fake claimer is unable to prove his ownership since his not aware of the correct key value.

Copy control is performed in a quite similar way. A digital image library administrator watermarks every digital image of the library with a constant well known key, before the content distribution takes place. This key is the declaration of the "never copy" instruction. Additionally, compliant devices are equipped with the detector of the watermarking mechanism along with the well know key. Upon the arrival of the watermarked digital image to the compliant device, the detector performs a watermarking detection. In case of positive response the compliant device understands the "never copy" instruction and forbids the replication of the image. This example illustrates the necessity of the device requirement to carry an incorporated detection mechanism, which is a quite ambitious expectation since the watermark detector essentially degrades the device functionality. Only law enforcement will make the above scenario appear as a realistic situation.

Consequently, the requirements of the copyright protection application of the digital library's security, are restricted to the casting of two watermarks, the first using the copyright owner's private key declaring his ownership, and the second using the well know constant key declaring the "never copy" instruction.

The next scenario concerns the digital signature security application and describes how the database administrator can discover an intruder trying to populate the database with malicious data. In digital image libraries of maximum importance and security the group of people authorized to contribute information is limited and well defined. The library administrator responsible for the validity of the content should maintain a record correlating a watermark key with the contributor's identification information. These keys are secretly distributed to the trusted party so as each authorized contributor should obtain a unique private key. When someone wishes to store information in the digital library, he sends the information along with his identification to the library administrator, only after he had watermarked the image using his private key. The image library administrator looks through his record and obtains the key related to the identity information provided by the unknown contributor. In case of positive response the administrator proceeds on storing the information to the library, in any other case the data are thrown away. In this way only the authorized group of people is permitted to contribute information to the digital library. The requirements of the digital signature security application are only one watermark per digital image and a Boolean response by the detector.

Finally the last scenario illustrates the transaction tracking security application, where the head of digital image library's security has the duty of tracking and capturing the information leak. As in the previous case this security application is applied in situations where the digital library content is very important and confidential. Once again the security administrator needs to maintain a record with numbers and names. The difference from the previous application is that now, no identity information is provided with the digital image, thus the security administrator has no way of knowing the correct key for the detection. The solution to this problem is the combination of a constant well-known key for the watermark casting, with a numerical detector's response allowing the correlation of the digital image with its original source. Just before the security administrator distributes the information to the authorized recipients, for example when a buyer is purchasing digital images, a watermark is embedded using the constant key. If a confidential image or document is found in the wrong hands, the security administrator can initialize the watermark detection process using always the constant key. The detector will result in a number indicating the original source of the image, likely responsible for the leak.

Summarizing the key requirements for the security purposes, every digital image included in the digital image library should contain a key for the owner identification application, a key for the copy control and a key representing the digital signature, all combined with a Boolean detector response. The last key requires a numerical output by the detector and refers to a transaction tracking facility applied mainly to e-commerce systems.

The proposed algorithm raises two basic issues concerning the watermarking technique. The first one is related with the data payload embedded into the image and the second with the detector ability to detect multiple watermarks. Data payload refers to the number of bits a watermark encodes within a unit of time or within a digital object. For a photograph, the data payload would refer to the number of bits encoded within the image.

The drawback in encoding a substantial number of bits into the image is the distortion introduced comparing to the original image. In our case the proposed watermark key administration requires three zero-bit watermarks (the detector's output is either one or zero) and one 14-bit watermark encoding 16384 different fingerprints. Mainly due to the inherent resilience of the DCT-domain technique [Barni, 98] the distortion introduced by the encoding of 17 bits is imperceptible as indicated by the calculated PSNR (Peak signal to noise ratio) value presented in the evaluation paragraph. By multiple watermarks we refer to the detector's potential of detecting a small amount of different watermarks into the same image without confusion. As in the previous case, the watermark algorithm's inherent capability solves the problem by maximizing the detector's output sufficiently above the selected threshold when the key is valid and minimizing it below the threshold in case of an irrelevant key value. The following graph (Fig. 8) demonstrates this feature.

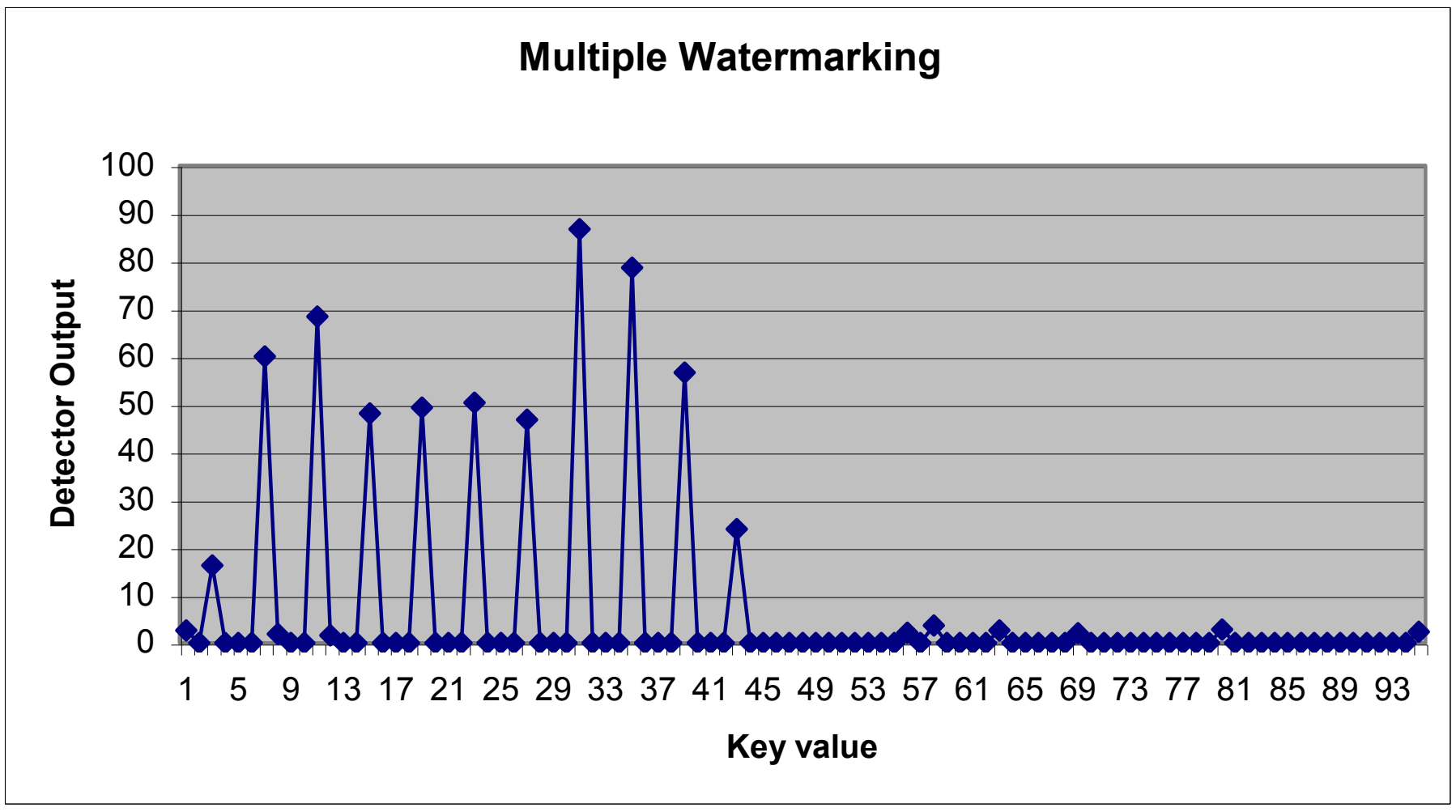

The detector used in the watermarking algorithm reveals the existence of 11 watermarks. Three of them correspond to the three zero-bit schemes while the rest 8 positive responses are used for the encoding of the fingerprint. The detector has succeeded in detecting all eleven watermarks without any confusion or misleading, resulting in a capability of facilitating proof of ownership, copy control, digital signature and transaction tracking at the same time.

## Conclusions

Most of the effort addressed in this work was dedicated on formulating a novel technique to embed robust multibit watermarks into digital images which also has an efficient and quick detection mechanism. The result was a technique applicable to every spread spectrum frequency domain watermarking method capable of hiding 214 different keys while maintaining a sufficient level of robustness. Special care was taken on resolving the potential problems derived from the process of casting multiple zero-bit watermarks onto the same coefficient area. Issues like the false positive probability, the image quality degradation and the robustness achieved by the proposed scheme were subject to thorough examination and evaluation.